\documentclass[twocolumn,showpacs,showkeys,amsmath,amssymb,pra,superscriptaddress]{revtex4-1}
\usepackage{epsfig}
\usepackage{epstopdf}
\usepackage{graphicx}
\usepackage{dcolumn}
\usepackage{color}
\usepackage{bm}
\usepackage{mathrsfs}
\usepackage{bbold}
\usepackage{amssymb}
\usepackage{upgreek}

\begin{document}

\title{Exact decoherence dynamics of $1/f$ noise}

\author{Md.~Manirul Ali}
\email{mani@mail.ncku.edu.tw}
\affiliation{Department of Physics, National Cheng Kung University, Tainan 70101, Taiwan}
\author{Ping-Yuan Lo}
\affiliation{Department of Physics, National Cheng Kung University, Tainan 70101, Taiwan}
\author{Wei-Min Zhang}
\email{wzhang@mail.ncku.edu.tw}
\affiliation{Department of Physics, National Cheng Kung University, Tainan 70101, Taiwan}

\date{\today}

\begin{abstract}
In this paper, we investigate the exact decoherence dynamics of a
superconducting resonator coupled to an electromagnetic reservoir characterized
by the $1/f$ noise at finite temperature, where a full
quantum description of the environment with $1/f^{x}$ noise (with $x \approx 1$)   is presented.
The exact master equation and the associated nonequilibrium Green's functions are solved exactly for
such an open system.
We show a clear signal of non-Markovian dynamics induced purely by $1/f$ noise.
Our analysis is  also applicable to another nano/micro mechanical oscillators. Finally, we demonstrate
the non-Markovian decoherence dynamics of photon number superposition states using Wigner distribution
that could be measured in experiments.
\end{abstract}


\maketitle

\section{Introduction}\label{sec:introduction}

Low frequency noise sprectrum $S(f) \sim 1/f$ was discovered in vacuum tubes and later observed
in a variety of systems \cite{systems1,systems2,systems3,systems4}. In electronics, this type of noise is commonly
referred to as $1/f$ noise. Miniaturization of any material system in nanometer-scale devices can further increase
$1/f$ noise levels and complicate practical applications \cite{mini,balandin}. The sensitivity of amplifiers and transducers
used in many types of sensor, particularly those that rely on an electrical response, is ultimately limited by the
low-frequency noise level \cite{sensor1,sensor2}. The importance of $1/f$ noise in electronics has motivated various studies
of its physical mechanisms and the development of a variety of methods for its reduction \cite{balandin}.
Different physical origin of fluctuation processes can be responsible for the $1/f$
noise in different materials and devices \cite{rmp1,rmp2,rmp3}.

In almost all quantum computing nanodevices, $1/f$ noise is found to be detrimental to the required maintenance of quantum coherent dynamics.
In the last two decades, several experiments with single electron transistors (SETs) \cite{SETn1,SETn2,SETn3,SETn4,SETn5}
and superconducting circuits \cite{SC1,SC2,SC3,SC4,SC5,Nori1,Nori2} at low temperatures revealed low-frequency noise with a spectrum
that scales inversely with frequency ($1/f$). The models and experimental observations \cite{fAlpha1,fAlpha2,fAlpha3,fAlpha4,fAlpha5,paris1,paris2}
suggest that the proper scaling of this low frequency noise should be $1/f^{x}$ where $x$ may differ slightly from unity. We will
investigate the exact decoherence dynamics of a superconducting resonator in presence of this noise with varying values of $x$,
including the situation under $1/f$ noise where $x \rightarrow 1$. Our analysis is also applicable to other nanomechanical
oscillators. Superconducting resonators and nano/micro mechanical oscillators have achieved sufficiently high
frequencies (GHz range), and work at very low temperature (mK range), in which nonclassical photon states can
be rather easily generated \cite{nonclass1,nonclass2,nonclass3}. Recently, $1/f$ noise is measured both in
superconducting resonator \cite{resonator1,resonator2} as well as in the mechanical oscillator \cite{oscillator1,oscillator2}.

Usually, in the literature on 1/f noise,  the environment is treated classically as a random field describing a stochastic process
\cite{fAlpha2,fAlpha3,fAlpha4,fAlpha5,paris1,paris2}. We investigate in this paper the exact decoherence dynamics of a
superconducting resonator coupled to an electromagnetic reservoir characterized by the $1/f$ frequency noise, where a full
quantum description of the environment with $1/f$ noise is presented. We also describe the decoherence dynamics of nonclassical
photon states of the resonator through the evolution of  Wigner distribution function. The paper is organized as follows.
In Sec.~\ref{sec:model}, we briefly discuss our model and the method we used. The noise power spectrum
is explored in Sec.~\ref{sec:noise} in the exact formalism of the nonequilibrium Green's functions. In Sec.~\ref{sec:results},
we present our numerical results on the exact decoherence dynamics of $1/f^{x}$ noise at finite temperature through the exact
solution of of the master equation associated with the nonequilibrium Green's functions. Using Wigner distribution,
we demonstrate in Sec.~\ref{sec:wigner}
the exact decoherence dynamics of superposition of photon number states under $1/f^{x}$ noise with different values
of $x$.  Finally, a conclusion is given in Sec.~\ref{sec:conclusion}.

\section{The model and the exact master equation}\label{sec:model}

We consider a superconducting or a nanomechanical resonator coupled
to an electromagnetic reservoir, the Hamiltonian of the total system can be written as
\begin{eqnarray}
\label{H}
H =  \hbar \omega_0 a^{\dagger} a + \sum_k \hbar \omega_k b_k^{\dagger} b_k
+ \sum_k \hbar \left( V_k a^{\dagger} b_k + V_k^{\ast} a b_k^{\dagger}  \right) ,
\end{eqnarray}
where the first term is the Hamiltonian $H_S$ of the single-mode resonator with frequency
$\omega_0$, and $a^{\dagger}$ and $a$ are the creation and annihilation operators of
the resonator quanta; the second term is the Hamiltonian $H_E$ of a general electromagnetic reservoir,
as a collection of infinite photon or phonon modes, where $b_k^{\dagger}$ and $b_k$ are the
corresponding creation and annihilation operators of the $k$-th photon or phonon mode with
frequency $\omega_k$. The third term is the system-reservoir coupling $H_I$ which characterizes
photon scattering processes with the scattering amplitude $V_k$ between the resonator and the
$k$-th reservoir mode. The nonlinear photonic processes have been ignored in Eq.~(\ref{H}) because the $1/f$ noise spectrum
only occurs in the very weak-coupling regime between the systems and its environment, as we will show
in Sec. III.

We shall use the exact master equation method to describe the decoherence dynamics of the resonator
under the influence of the $1/f$ noise. The master equation is given in terms of the reduced density
operator, obtained from the density operator of the total system by tracing
over the environmental degrees of freedom $\rho(t)=Tr_E[\rho_{tot}(t)]$.
The total density operator is governed by the quantum evolution operator:
$\rho_{tot}(t) = e^{-\frac{i}{\hbar}H(t-t_0)} \rho_{tot}(t_0) e^{\frac{i}{\hbar}H(t-t_0)}$.
As it was originally proposed by Feynman and Vernon \cite{influence1,influence2,influence3,influence4},
we take the initial state of the total system as a directly
product of an arbitrary initial state of the system with the thermal state of the reservoir,
$\rho_{tot}(t_0)=\rho(t_0) \otimes \rho_E(t_0)$, where $\rho_E(t_0)= \exp(-\beta H_E)
/Tr[\exp(-\beta H_E)]$. In this case, tracing over all the environmental degrees of
freedom can be easily carried out using the Feynman-Vernon influence functional
approach \cite{influence1,influence2} in the framework of coherent-state path-integral
representation \cite{Zhang90,annphys}. The resulting master equation
for the reduced density operator has the following form \cite{annphys}:
\begin{eqnarray}
\label{master}
\frac{d\rho(t)}{dt} &=& -i \omega_0^{\prime} (t) \left[ a^{\dagger} a , \rho(t) \right] \\
\nonumber
&&{} + \gamma(t) \left[ 2 a \rho(t) a^{\dagger} - a^{\dagger} a \rho(t) - \rho(t) a^{\dagger} a \right] \\
\nonumber
&&{} + \widetilde{\gamma}(t) \left[ a \rho(t) a^{\dagger} + a^{\dagger} \rho(t) a - a^{\dagger} a \rho(t)
- \rho(t) a a^{\dagger} \right] ,
\end{eqnarray}
where the time-dependent coefficient $\omega_0^{\prime} (t)$ is the renormalized frequency of the resonator,
while $\gamma(t)$ and $\widetilde{\gamma}(t)$ describe the dissipation (damping) and fluctuation (noise) of
the resonator due to its coupling to the reservoir. These coefficients can be exactly determined by the
following relations \cite{annphys,bosonic1a,bosonic1b,bosonic2}
\begin{eqnarray}
\nonumber
\omega_0^{\prime}(t) &=& -Im\left[\frac{{\dot u}(t)}{u(t)}\right], ~~
\gamma(t) = -Re\left[ \frac{{\dot u}(t)}{u(t)} \right], \\
\widetilde{\gamma}(t) &=& {\dot v}(t) - 2 v(t) Re\left[ \frac{{\dot u}(t)}{u(t)} \right].
\end{eqnarray}
The function $u(t)$ is the nonequilibrium propagating (or spectral) Green's function
of the system, which satisfies the Dyson equation of motion
\begin{eqnarray}
{\dot u}(t) + i \omega_0 u(t) + \int_{t_0}^t d\tau g(t-\tau) u(\tau) = 0,
\label{ide}
\end{eqnarray}
subject to the initial condition $u(t_0)=1$. The nonequilibrium thermal fluctuation is characterised by
the correlation function $v(t)$ through the nonequilibrium fluctuation-dissipation theorem \cite{general},
which is given explicitely by
\begin{eqnarray}
v(t) = \int_{t_0}^t d\tau \int_{t_0}^t d\tau^{\prime}~ {\tilde g}(\tau-\tau^{\prime}) u^{\ast}(\tau) u(\tau^{\prime}) .
\label{vt}
\end{eqnarray}
The time correlation functions
$g(t-\tau)$ and ${\tilde g}(\tau-\tau^{\prime})$ in Eqs.~(\ref{ide}) and (\ref{vt}) are given by
\begin{eqnarray}
g(t-\tau) = \int_0^{\infty} \frac{d\omega}{2\pi} J(\omega) e^{-i\omega(t-\tau)}
\label{tcf1}
\end{eqnarray}
\begin{eqnarray}
{\tilde g}(\tau-\tau^{\prime}) = \int_0^{\infty} \frac{d\omega}{2\pi} J(\omega) {\bar n}(\omega,T)
e^{-i\omega(\tau-\tau^{\prime})}
\label{tcf2}
\end{eqnarray}
which characterize all the non-Markovian back-action memory effects between the system and the reservoir,
where $J(\omega)=2\pi\sum_k |V_k|^2 \delta(\omega-\omega_k)$ is the spectral density, and $V_k$ is
the coupling between the system and the reservoir. Furthermore, ${\bar n}(\omega,T)=\frac{1}{e^{\hbar \omega / k_B T}-1}$
is the particle number distribution of the bosonic reservoir at the initial temperature $T$. If the reservoir spectrum is continuous,
$V_k \rightarrow V(\omega)$, we have $J(\omega)=2\pi \varrho(\omega) |V(\omega)|^2$ where $\varrho(\omega)$ is the density
of state of the reservoir.

\section{Quantum description of $1/f$ Noise spectrum}\label{sec:noise}

In the literature on 1/f noise, the environment is treated classically \cite{fAlpha2,fAlpha3,fAlpha4,fAlpha5,paris1,paris2} with a classical random
field $c(t)$ describing a stochastic process. The system-environment coupling is described by a term $\sim c(t) A$, and $A$ is a system
operator.
Then the system Hamiltonian becomes stochastic due to
the random nature of the classical random field $c(t)$ which can take different form corresponding to different kinds of classical noise.
A typical classical noise source
is the random telegraphic noise (RTN) which is used to classically model an environment for solid-state devices \cite{rmp3,paris1,paris2} (and references therein),
where the system is considered to be interacting with a bistable fluctuator for which the time-dependent parameter randomly flipping between two
values $c(t)=\pm 1$ with a switching rate $\nu$. The Hamiltonian with such a noise source describeing a system subject to a random telegraph noise
is usually used as a basic building block to describe noises of the type $1/f^{\alpha}$, and is characterized by an exponentially decaying correlation
function $e^{-\nu |t|}$ of the fluctuating quantity $c(t)$. Then the noise spectrum is a Lorentzian function
\begin{eqnarray}
\nonumber
{\cal S}(\omega,\nu) &=& \int_{-\infty}^{+\infty} dt e^{i\omega t}~ \overline{c(t) c(0)}  \\
&=& \int_{-\infty}^{+\infty} dt e^{i\omega t} e^{-\nu |t|} = \frac{1}{\pi} \frac{\nu}{\omega^2 + \nu^2} .
\label{rtn1}
\end{eqnarray}
To reproduce the $1/f^{\alpha}$ spectrum, the single RTN frequency power spectrum is then integrated over the switching rates $\nu$ with
a suitable probability distribution:
\begin{eqnarray}
{\cal S}_{1/f^{\alpha}}(\omega) = \int_{\nu_1}^{\nu_2} {\cal S}(\omega,\nu) p_{\alpha}(\nu) d\nu,
\label{rtn2}
\end{eqnarray}
The integration is generally performed between a minimum and a maximum value of the switching rates, $\nu_1$ and $\nu_2$ respectively.
In order to simulate $1/f^{\alpha}$ noise spectrum, the switching rate distribution $p_{\alpha}(\nu)$ is considered to be proportional
to $1/\nu^{\alpha}$. When the integration in Eq.(\ref{rtn2}) is performed, the spectrum shows $1/f^{\alpha}$ behavior in a frequency
interval, so that all frequencies belonging to such an interval satisfies the condition $\nu_1 \le f \le \nu_2$. Although this low and high
frequency cutoff frequencies are artificially fixed in the literature \cite{SC1,SC2,SC3,SC5,Cutoff1}, the physical origin of the cutoff
frequencies are debatable \cite{Cutoff2}. Another crucial point is that the classical environment with the noise spectrum $1/f^{\alpha}$
can be realized by different configuration of bistable fluctuators.  Noise spectrum $1/f^{\alpha}$ can be obtained either considering a
single bistable fluctuator whose switching rate is randomly chosen from a distribution $p_{\alpha}(\nu)$ as shown above. The same
$1/f^{\alpha}$ spectrum can also be realized from the coupling of a system with a large number of fluctuators, where the noise spectrum
can be obtained as a result of linear combination of many Lorentzian, each characterized by a specific switching rate
\cite{rmp3,paris1,paris2}. As already pointed out in several papers \cite{paris1,nine1,nine2,nine3,nine4}, different microscopic configuration of
the environment leading to the same spectra may correspond to different physical phenomena. Thus, mere knowledge of the noise spectrum is not
sufficient to describe the environmental influence on the quantum dynamics of open systems and it is necessary to specify the model for the noise
source in more detail.

In the previous section, we have presented a full quantum-mechanical description of the
system-environment coupling for a superconducting resonator coupled to an electromagnetic
reservoir. Before we explore the decoherence dynamics of a superconducting resonator or
a nanomechanical resonator, induced by the $1/f$ noise, it is important to justify
the conditions for the occurrence of $1/f$ noise in a given electromagnetic reservoir
characterized by the spectral density $J(\omega)$. In the literature, the spectral density
$J(\omega)$ for an electromagnetic reservoir is found to be Ohmic-type \cite{influence3},
given by
\begin{eqnarray}
J(\omega) = 2\pi \eta ~\omega \left( \frac{\omega}{\omega_c} \right)^{s-1} ~e^{-\omega/\omega_c},
\label{sd}
\end{eqnarray}
with $0<x=(1-s)<1$. It has been pointed out \cite{schon} that the $1/f$ noise spectrum corresponds to the special case of the spectral
density (\ref{sd}) with $s \approx 0$ or $x\approx1$. However, the exact connection of $J(\omega)$
with the noise spectrum $S(\omega)$ has not been carried out so far in the literature. Here, within the
exact master equation formalism, we find that the exact solutions of the nonequilibrium
Green's functions $u(t)$ and $v(t)$ carry all the information on the quantum nature of
the noise spectrum $S(\omega)$.

Specifically, the noise spectrum of the system is quantum mechanically defined by the Fourier
transform of the two-time particle correlation function:
\begin{eqnarray}
S(\omega) = \lim_{ t \rightarrow \infty} \int_{-\infty}^{\infty} e^{i\omega \tau}
\langle a^{\dagger} (t+\tau) a(t) \rangle ~d\tau .
\label{noise1}
\end{eqnarray}
As we have shown recently \cite{annphys},   the two-time particle correlation function
$\langle a^{\dagger} (t+\tau) a(t) \rangle$ obeys the following relation in our exact master equation
formalism,
\begin{eqnarray}
\nonumber
\langle a^{\dagger} (t+\tau) a(t) \rangle = u^\ast(t+\tau)u(t) \langle a^{\dagger}(t_0) a(t_0) \rangle
+ v(t,t+\tau) ,
\end{eqnarray}
where
\begin{eqnarray}
\nonumber
v(t,t+\tau) = \int_{t_0}^t d\tau_1 \int_{t_0}^{t+\tau} d\tau_2 u(t,\tau_1) {\tilde g}(\tau_1,\tau_2) u^{\ast}(t+\tau,\tau_2), \\
\label{vtwo}
\end{eqnarray}
which is the more general form of Eq.~(\ref{vt}). On the other hand, the exact analytic solution of the integro-differential
equation (\ref{ide}) is also recently given in \cite{general},
\begin{eqnarray}
\nonumber
u(t) = \mathcal{Z} e^{-i \omega_b (t-t_0)} + \frac{1}{\pi} \int_0^\infty
\frac{\gamma(\omega) e^{-i\omega(t-t_0)}}{\left[ \omega - \omega_0 -
\Delta(\omega) \right]^2 + \gamma^2(\omega)} d\omega , \\
\label{ut}
\end{eqnarray}
where $\gamma(\omega)=J(\omega)/2$ and $\Delta(\omega) = {\cal P}\int_0^\infty \frac{J(\omega')}{\omega-\omega'} d\omega' $,
which are the real and imaginary parts of the self-energy correction, $\Sigma(z)$, to the resonator, induced by the coupling
between the resonator and the environment,
\begin{eqnarray}
\Sigma(z) = \int_0^\infty \frac{J(\omega)}{z-\omega} d\omega .
\label{sigmaz}
\end{eqnarray}
The first term in Eq.~(\ref{ut}) is the contribution of the dissipationless localized
mode, where the localized mode frequency $\omega_b$ is determined by $\omega_b - \omega_0 - \Delta(\omega_b)=0$, and
$\mathcal{Z}=\left[ 1 - \Sigma^{\prime}(\omega_b) \right]^{-1}$ corresponds to the residue of $\Sigma(z)$ at the pole
$z=\omega_b$, which gives the amplitude of the localized mode.

\begin{figure}[h]
\centering{\rotatebox{0}{\resizebox{8.5cm}{5.0cm}{\includegraphics{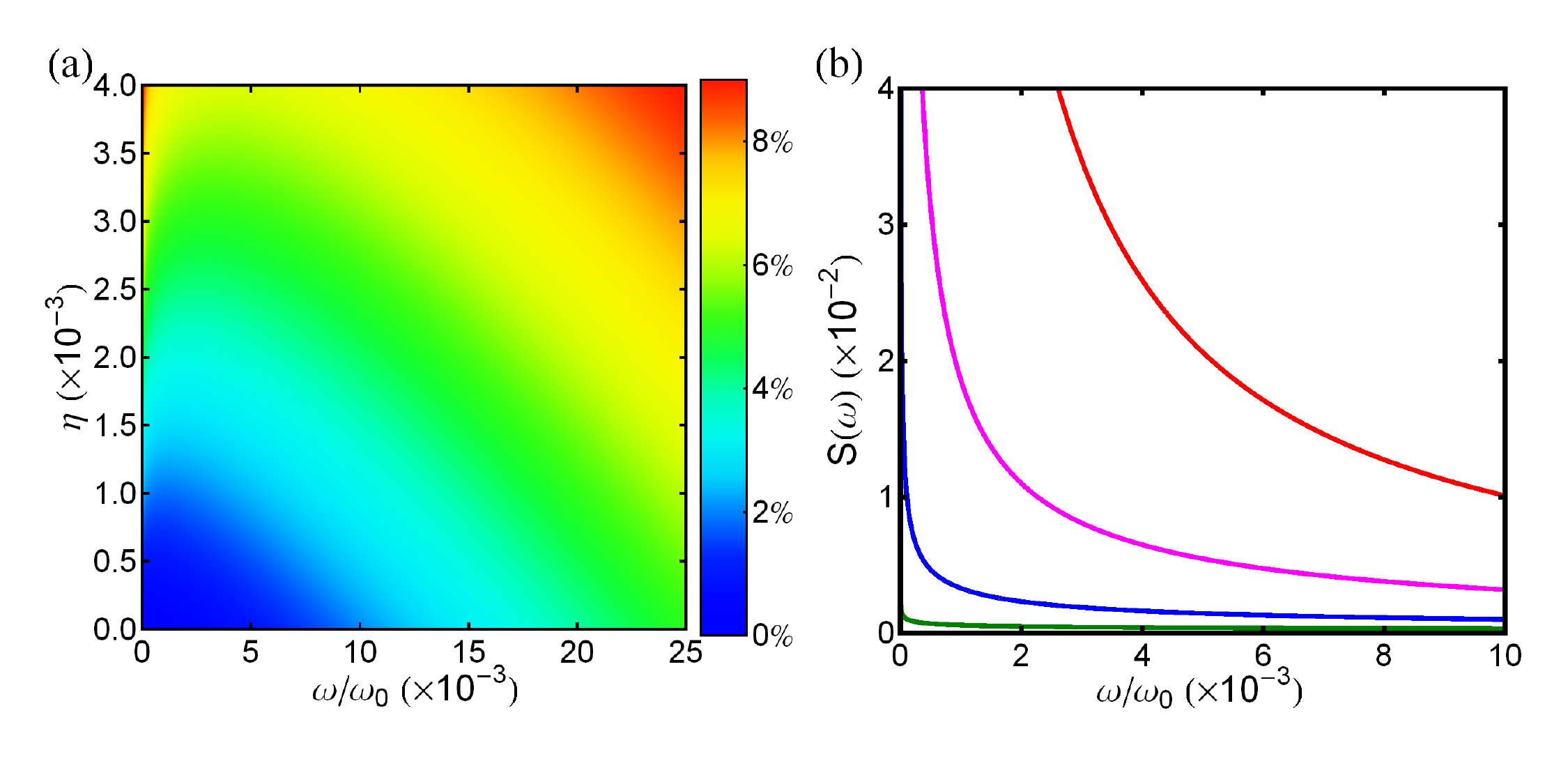}}}}
\caption{\label{noiserange}  (Color online) (a) We plot the first order correction term $2\xi \zeta$ in the expansion of the noise spectrum
(\ref{correc}) for $\omega_c = \omega_0$, which shows a very narrow range of $\eta$ and $\omega$ for the low frequency noise spectrum
behaving as $1/f^x$ power law. (b) The noise spectrum $S(\omega)$ is plotted for different values of $x$ as
$x = 0.25$ (green), $x = 0.5$ (blue), $x = 0.75$ (pink) and $x = 0.9999$ (red) at $\eta=10^{-3}$
with $\omega_c = \omega_0 = 5$ GHz, and $T=25$ mK.}
\end{figure}

Using the exact solution of Eq.(\ref{ut}) and the relation (\ref{vtwo}), we find that the noise spectrum is given by
\begin{align}
S(\omega) &= {\mathcal{Z}}^2 \delta(\omega-\omega_b) \langle a^{\dagger}(t_0) a(t_0) \rangle
+ \Bigg[ \frac{{\mathcal{Z}}^2 J(\omega)}{(\omega-\omega_b)^2}
\nonumber \\
\nonumber
& +
\frac{J(\omega)}{\left[ \omega - \omega_0 - \Delta(\omega) \right]^2 + \gamma^2(\omega)} \Bigg] {\bar n}(\omega,T) \\
& = S_1(\omega) + S_2(\omega),
\end{align}
where $S_1(\omega)$ and $S_2(\omega)$ are the contributions of the particle correlations from the system and the environment respectively
to the noise spectrum, due to the coupling between them. Now, $S_2(\omega)$ can have a power series expansion with Legendre polynomials
$P_n(\xi)$ as
\begin{align}
\nonumber
S_2(\omega) & = \frac{{\mathcal{Z}}^2 J(\omega){\bar n}(\omega,T)}{(\omega-\omega_b)^2} +  \frac{J(\omega){\bar n}(\omega,T) }{\omega_0^2 + \gamma^2(\omega)}
\Big( \sum_{n=0} P_n(\xi) \zeta^n \Big)^2  \\
\nonumber
&= \frac{{\mathcal{Z}}^2 J(\omega){\bar n}(\omega,T)}{(\omega-\omega_b)^2}
+ \frac{J(\omega){\bar n}(\omega,T)}{\omega_0^2 + \gamma^2(\omega)} \Bigg[ 1 + 2 \xi \zeta
\\
& + \mathcal{O}(\xi^2 \zeta^2) + ...\Bigg],
\label{correc}
\end{align}
where $\xi = \omega_0/\sqrt{\omega_0^2 + \gamma^2(\omega)}$, and $\zeta =
 (\omega -\Delta(\omega))/\sqrt{\omega_0^2 + \gamma^2(\omega)}$.
We should only take the low-frequency limits ${\hbar \omega}/{k_B T} \ll 1$ and $\omega \ll \omega_0$,
in order to see the low frequency behavior of the noise spectrum.
Through a numerical check, see figure~\ref{noiserange}(a), we find that for $\eta \le 10^{-3}$
and $\omega/\omega_0 \le 10^{-2}$, one has $\xi\zeta \ll 1$ and
$\omega_0^2 \gg \gamma^2(\omega)$. Meanwhile, when the coupling between the system
and the environment $\eta \le 10^{-3}$, the localized mode
amplitude (${\mathcal{Z}}$) becomes negligibly small, and therefore it does not play any role for
low frequency behavior of the noise spectrum. Thus the noise
spectrum is reduced to
\begin{eqnarray}
S(\omega) = \frac{\eta^\prime k_B T}{ \omega^{1-s} } = \frac{\eta^\prime k_B T}{ \omega^x },
\label{noiseF}
\end{eqnarray}
where $\eta^{\prime}=\eta \omega_c^{1-s}/\hbar \omega_0^2$. For $s=0$, Eq.(\ref{noiseF})
gives the exact $1/f$ noise spectrum. Generally, the
low frequency noise ($1/f^x$) dominates when $x > 0$ or $s < 1$
in the very weak coupling regime between the system and its reservoir.
In figure~\ref{noiserange}(a), we show the range of $\eta$ and $\omega$
where the low frequency noise spectrum behaves as $1/f^x$ power law.
For $x\approx1$, the $1/f$ noise behavior shows up in the low frequency domain
where the coupling strength $\eta$ must be sufficiently weak. We also plot in
figure~\ref{noiserange}(b) the noise spectrum $S(\omega)$ for
$\eta=10^{-3}$ with different values of $x$. Different $1/f^x$ behaviors are displayed.

The above noise power spectrum, calculated analytically and exactly through the
Fourier transform of the two-time particle correlation function (\ref{noise1}), provides
a fully quantum mechanical description of the $1/f^x$ noise. This quantum mechanical description
of the $1/f$ noise shows that it valids in a very narrow range of $\eta$ and $\omega$ when both
the frequency and the coupling strengths are very small, in comparison with the energy scale of the system.

\section{Decoherence dynamics under $1/f^x$ noise}\label{sec:results}

\begin{figure}[h]
\centering{\rotatebox{0}{\resizebox{8.5cm}{5.0cm}{\includegraphics{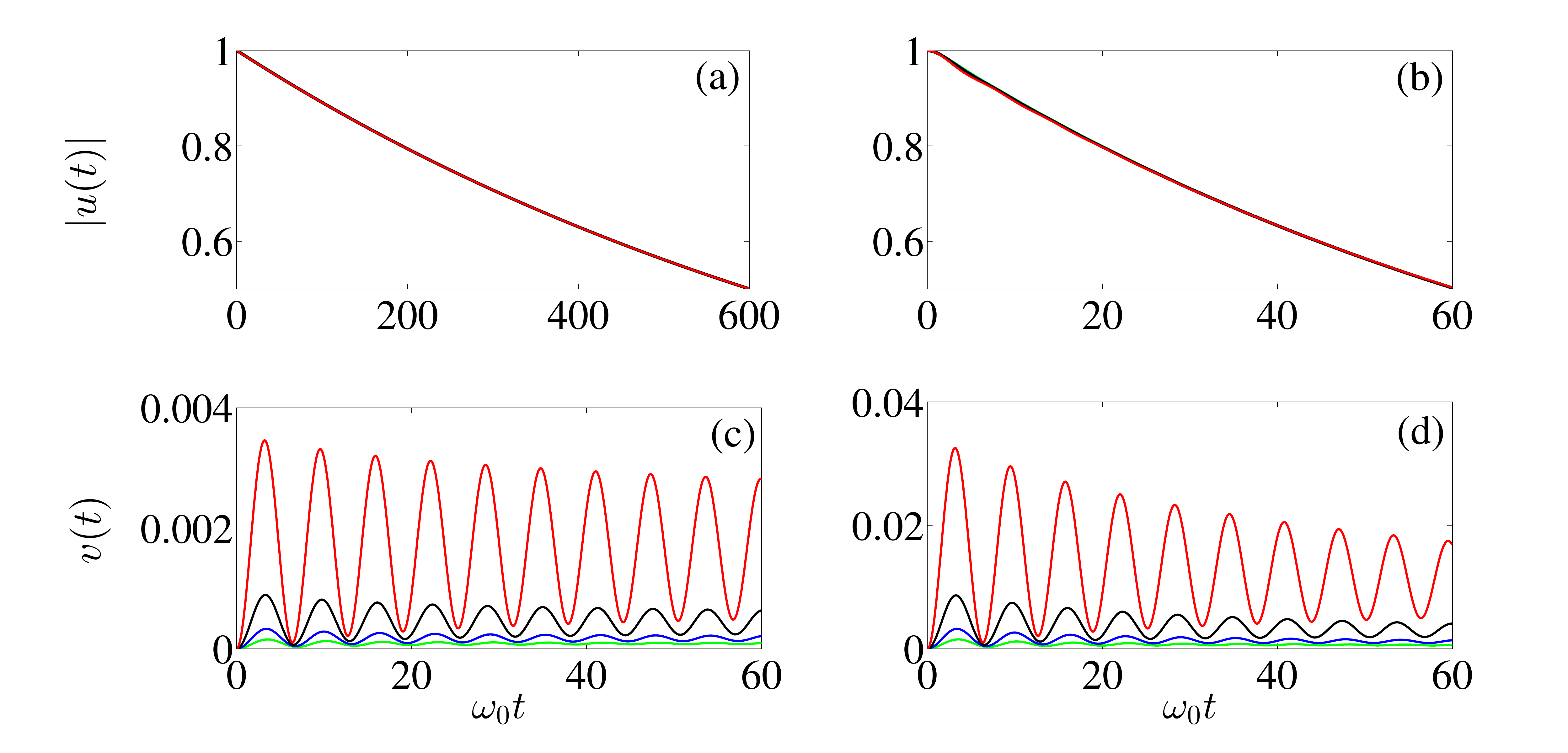}}}}
\caption{\label{uv}  (Color online) Comparison of the exact solution of $|u(t)|$ and $v(t)$
for $1/f^x$ noise with $x=0.25$ (green), $x=0.5$ (blue), $x=0.75$
(black), $x=0.9999$ (red). We plot the exact $|u(t)|$ and $v(t)$ for different
system-environment coupling $\eta$: (a,c) $\eta=10^{-3}$ and (b,d) $\eta=10^{-2}$. The other
parameters are taken as $\omega_c = \omega_0 = 5$~GHz, and $T=25$~mK.}
\end{figure}

Now we shall study the exact decoherence dynamics induced by $1/f$ noise under the condition
${\hbar \omega}/{k_B T} \ll 1$ and $\omega \ll \omega_0$, and the dimensionless coupling
strength $\eta$ is sufficiently weak. We start with the $1/f^x$ noise with $x=0.25$,
and gradually increase it to $x=0.5,~0.75$, and $0.9999$, to examine the change of dissipation and
fluctuation dynamics of the resonator when the noise spectrum approaches to $1/f$.
Dissipation and fluctuation dynamics through the exact solution of $u(t)$ and $v(t)$ are presented in Figure~\ref{uv}
for $1/f^x$ noise (\ref{sd}) with $x=0.25,~0.5,~0.75,$ and $0.9999$, respectively, corresponding
to the four different curves in each graph. It shows how dissipation and fluctuation change as the reservoir spectra
approach to low-frequency-dominated regime. One can see that for very weak coupling ($\eta=10^{-3}$), the particle
propagating function $u(t)$ has similar damping dynamics (monotonous decay) with different $x$ values, see figure \ref{uv}(a).
We see similar decay dynamics for $u(t)$ (without having any non-Markovian effect) even if we increase the coupling
strength to $\eta=10^{-2}$ with changing $x$ values, see figure \ref{uv}(b). Figures \ref{uv}(c) and (d) show thermal fluctuations
in terms of the correlation Green's function $v(t)$. The correlation Green's
function $v(t)$ quantifies physically the thermal-fluctuation-induced average particle number inside the resonator, reflecting the
noise effect of the decoherence dynamics. Figure \ref{uv}(c) shows that the decoherence dynamics is significantly
different for $1/f$ spectrum ($x=0.9999$), as the oscillations of $v(t)$ is very strong, compared to other $1/f^x$
(with $x=0.25,~0.5,~0.75$) at weak coupling ($\eta=10^{-3}$). This is the signal of a non-Markovian decoherence dynamics purely induced
by the noise effect.
Physically this can be seen from Eq.~(\ref{tcf2}) that the initial particle distribution function $\overline{n}(\omega,T)$ induces
explicitly frequency dependence to the memory kernel $\widetilde{g}(\tau-\tau')$, and in particular, this frequency dependence
becomes stronger in the low frequency regime $\sim 1/\omega$.
This distinct oscillatory feature of $v(t)$ persists even at higher
coupling strength ($\eta=10^{-2}$), although it has a long time decay behavior, see figure~\ref{uv}(d).
In conclusion, we show here, for the first time, the physical mechanism
of non-Markovian dynamics induced by $1/f$ noise even though the system-environment
coupling is very small.
\begin{figure}[h]
\centering{\rotatebox{0}{\resizebox{8.5cm}{5.0cm}{\includegraphics{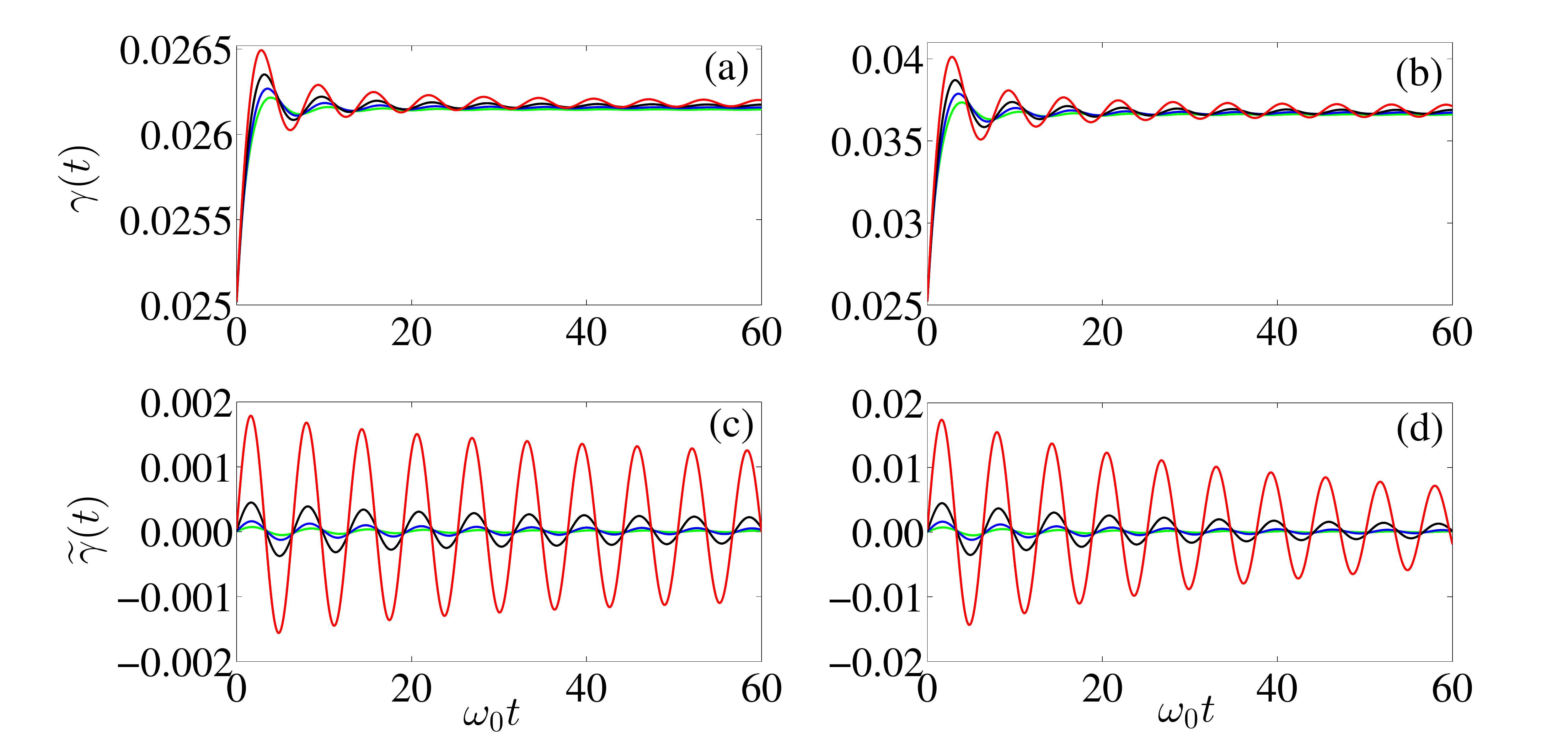}}}}
\caption{\label{Pg12}  (Color online) The dissipation and fluctuation coefficients $\gamma(t)$ and
$\widetilde{\gamma}(t)$ are calculated from the exact solution of $u(t)$ and $v(t)$ for $1/f^x$ noise
with $x=0.25$ (green), $x=0.5$ (blue), $x=0.75$ (black), $x=0.9999$ (red).
We plot exact $\gamma(t)$ for different system-environment coupling $\eta$ (a) $\eta=10^{-3}$ and (b) $\eta=10^{-2}$.
We next plot exact $\widetilde{\gamma}(t)$ at (c) $\eta=10^{-3}$ and (d) $\eta=10^{-2}$. The values of other
parameters are $\omega_c = \omega_0 = 5$~GHz, and $T=25$~mK.}
\end{figure}

\begin{figure}[h]
\centering{\rotatebox{0}{\resizebox{8.0cm}{4.0cm}{\includegraphics{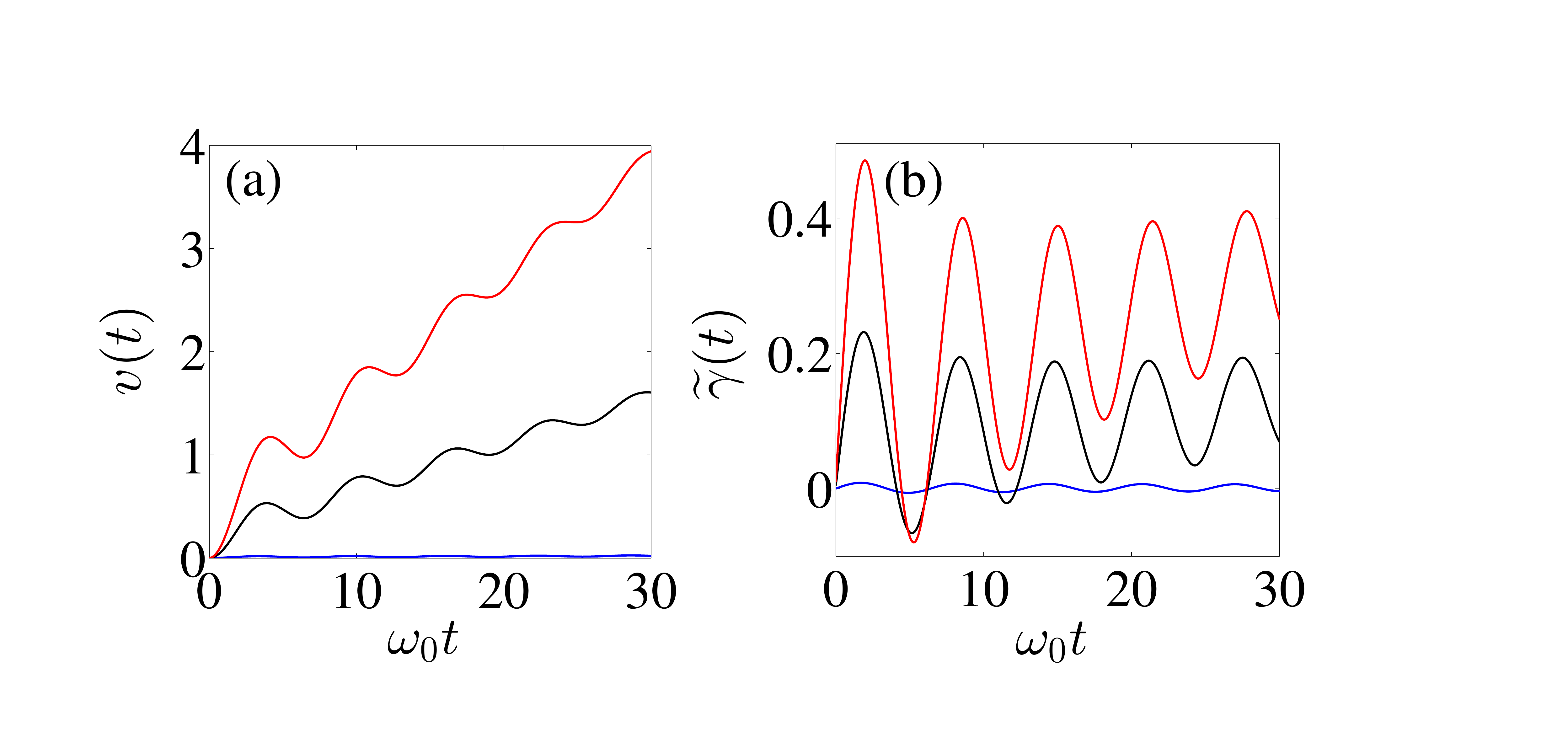}}}}
\caption{\label{vg2} (Color online) (a) the exact $v(t)$ and (b) the fluctuation coefficient $\widetilde{\gamma}(t)$
(calculated from the exact solution of $u(t)$ and $v(t)$) for $1/f$ spectrum with $x=0.9999$ at various
temperatures $T=25$~mK (blue), $T=1.0$~K (black), and $T=2.5$~K (red). The
other parameters are $\eta=10^{-3}$, and $\omega_c = \omega_0 = 1$~GHz.}
\end{figure}

With the above exact solution of the dissipation and fluctuation dynamics, we now present the dissipation and fluctuation
coefficients in the master equation, $\gamma(t)$ and $\widetilde{\gamma}(t)$, which manifests quantitatively
the decoherence behavior of $1/f$ noise.  Figures \ref{Pg12}(a) and (b) show the time evolution of the dissipation
coefficient $\gamma(t)$ for different $1/f^x$ noise with $x=0.25,~0.5,~0.75,$ and $0.9999$ at two different
values of the coupling strengths, $\eta=10^{-3}$ and $10^{-2}$, partnered to the solutions in figures \ref{uv}(a) and (b).
In the very weak coupling region ($\eta=10^{-3}$), including the $1/f$ noise, we see from figure~\ref{Pg12}(a) that the
dissipation coefficient $\gamma(t) >0$, which indicates that the corresponding dissipation solution is always Markovian
\cite{PRL10a,PRL10b}. A similar behavior of $\gamma(t)$ with increased magnitude is seen when the coupling
strength is increased ($\eta=10^{-2}$ for figure~\ref{Pg12}(b)). However, the fluctuation coefficient $\widetilde{\gamma}(t)$
is very sensitive to the noise spectrum and it behaves qualitatively different from the dissipation coefficient $\gamma(t)$ in the
weak-coupling regime, in particular, in the $1/f$ noise regime. As one can see from figure~\ref{Pg12}(c), $\widetilde{\gamma}(t)$
oscillates between positive and negative values resulting to a non-Markovian memory effect. This positive and negative bounded
value is significantly larger in case of $1/f$ noise spectrum ($x \sim 1$), compared to other spectrums with smaller values of $x$.
The distinct oscillatory feature of $\widetilde{\gamma}(t)$ persists even at higher coupling strength ($\eta=10^{-2}$), with a long
time decay behavior, see figure \ref{Pg12}(d). The presence of this persistent oscillation between positive and negative values in
fluctuation coefficient $\widetilde{\gamma}(t)$ but not in the dissipation coefficient $\gamma(t)$, shows the evidence for strong non-Markovian
dynamics, associated with the $1/f$ noise, as we have just pointed out in the analysis of the correlation Green's function $v(t)$.
The decoherence dynamics of the resonator is very sensitive to the temperature for $1/f^x$ spectrum. To show the effect of
temperature dependence on the decoherence dynamics, we plot the non-equilibrium thermal fluctuation $v(t)$ and the fluctuation coefficient
$\widetilde{\gamma}(t)$ in figure~\ref{vg2} for $1/f^x$ spectrum with $x=0.9999$ at various temperatures
$T=25$~mK, $T=1.0$~K, and $T=2.5$~K, respectively.  It shows that the
magnitude of both $v(t)$ and $\widetilde{\gamma}(t)$ becomes significant with the rising temperature even if the
system-reservoir coupling is very small ($\eta=10^{-3}$). This temperature dependence
comes through the particle number distribution ${\bar n}(\omega,T)$ in Eq.(\ref{tcf2}), as purely a noise effect.

\section{Decoherence dynamics of the resonator under $1/f^x$ noise using Wigner distribution}\label{sec:wigner}

Next, we explore the decoherence dynamics of quantum photon states under the $1/f$ noise by
examining the evolution of the corresponding Wigner function. With the help of the exact master equation
(\ref{master}), the exact Wigner function of an arbitrary quantum state at arbitrary time $t$ in the complex
space $\{z\}$ is given by
\begin{eqnarray}
\nonumber
W(z,t) = \int d\mu(\alpha_0) d\mu(\alpha_0^\prime) \langle \alpha_0|\rho(t_0)|\alpha_0^\prime \rangle
{\mathscr T}(z,t|\alpha_0,\alpha_0^{\prime \ast},t_0), \\
\label{wigner}
\end{eqnarray}
where $|\alpha\rangle=e^{\alpha a^{\dagger}} |0\rangle$ is the coherent state,
$d\mu(\alpha)=\frac{d\alpha^{\ast} d\alpha}{2\pi i} e^{-|\alpha|^2}$ is the integral measure of the Bergman
complex space, $\rho(t_0)$ is the reduced density matrix of the initial state, and the propagating function
${\mathscr T}(z,t|\alpha_0,\alpha_0^{\prime \ast},t_0)$ is given by \cite{bosonic2, Rancon}
\begin{eqnarray}
{\mathscr T}(z,t|& \alpha_0, \alpha_0^{\prime \ast},t_0) = W_{0}^{0} (z,t) \exp\{ z^{\ast} \Omega(t) u(t) \alpha_0
\nonumber  \\
& + z \Omega(t) u^{\ast}(t) \alpha_0^{\prime \ast} + \alpha_0^{\prime \ast} \left[ 1 - |u(t)|^2 \Omega(t) \right] \alpha_0  \} ,
\label{pgtr}
\end{eqnarray}
where
\begin{eqnarray}
\nonumber
\Omega(t) = \frac{2}{1 + 2 v(t)} ~~ \mbox{and} ~~ W_{0}^{0} (z,t) = \frac{2 \exp\left( - \Omega(t) |z|^2  \right)}{\pi \left[ 1 + v(t) \right]}.
\end{eqnarray}

\begin{figure}[h]
\centering{\rotatebox{0}{\resizebox{8.0cm}{6.0cm}{\includegraphics{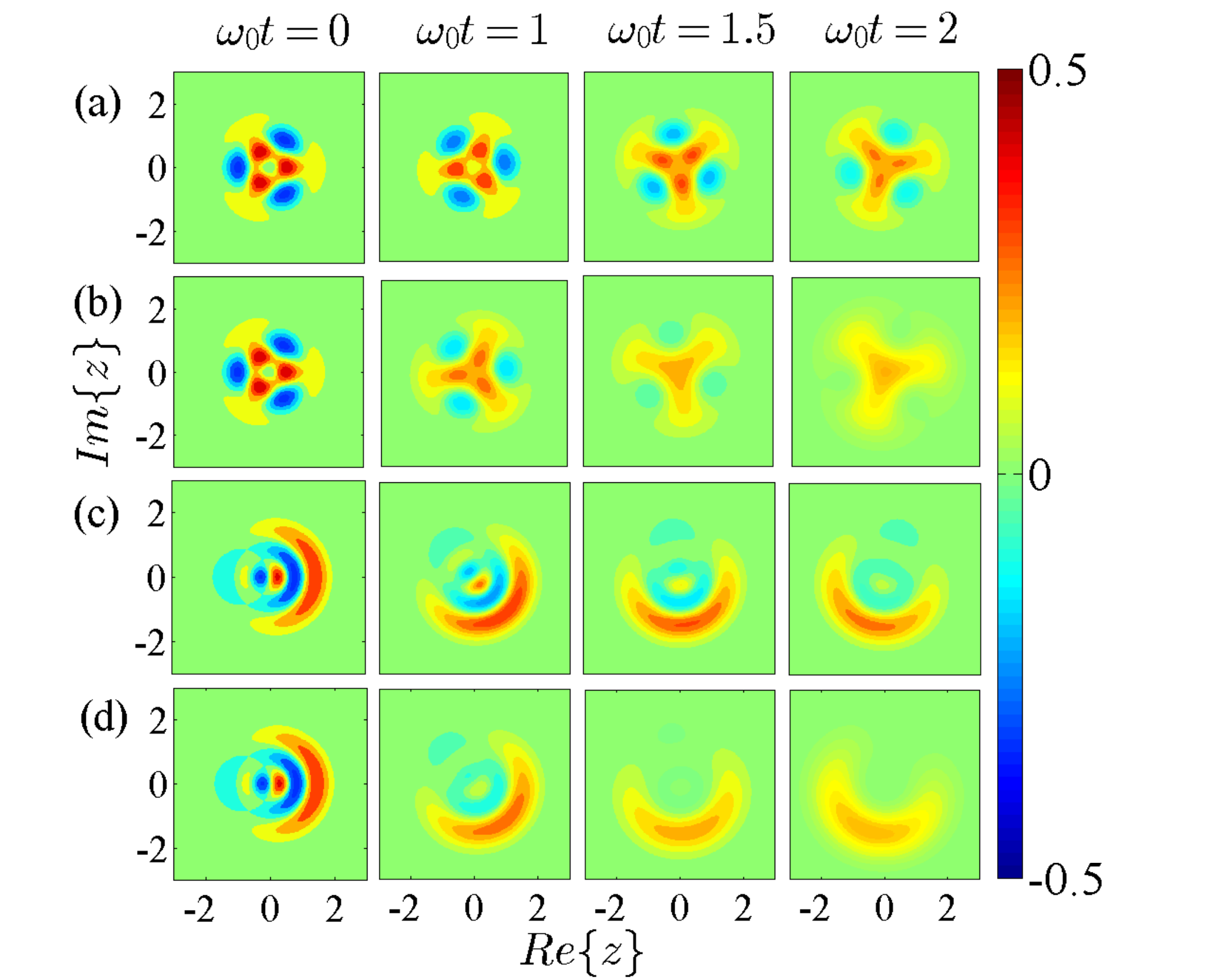}}}}
\caption{\label{Super2}  (Color online) The decoherence dynamics of the resonator (prepared in initial
superposition states $1/\sqrt{2}\left( |0\rangle + |3\rangle \right)$ and $1/\sqrt{2}\left( |2\rangle
+ |3\rangle \right)$ are respectively shown by plotting the time evolution of the Wigner function
for $1/f^x$ noise with (a) and (c) with $x=0.25$, and also (b) and (d) with $x=0.9999$.
The values of other parameters are $\eta=10^{-3}$,~ $\omega_c = \omega_0 = 1$~GHz, and
$T=2.5$~K.}
\end{figure}

We investigate the damping and decoherence dynamics of the resonator in the presence of low frequency
$1/f^x$ noise at a finite temperature. We see the effect of changing values of $x$ on the dynamics of the
resonator which is prepared initially in a superposition of Fock states. If the resonator is initially prepared
in a photon number superposition state $\rho(t_0)=|\psi_0\rangle\langle\psi_0|$ with
$|\psi_0\rangle=1/\sqrt{2}\left( |0\rangle + |n\rangle \right)$. The time evolved Wigner function in this
case is given by
\begin{eqnarray}
W_{0}^{n} (z,t) &=& \frac{1}{2} \left[ W_{0}^{0} (z,t) + W_{n}^{n} (z,t) \right] \\
\nonumber
&+& \frac{W_{0}^{0} (z,t)}{2\sqrt{n!}} \left[ (z \Omega(t) u^{\ast}(t))^n + (z^{\ast} \Omega(t) u(t))^n  \right] ,
\end{eqnarray}
where $W_{0}^{0} (z,t)$ is the time evolved Wigner function for the initial vacuum state, and
\begin{eqnarray}
W_{n}^{n} (z,t) &=& W_{0}^{0} (z,t) \sum_{p=0}^n \frac{n!}{p!(n-p)!(n-p)!} \times \\
\nonumber
&&{} \left[ \left(|u(t)|^2~\Omega^2(t)~|z|^2 \right)^{n-p}
\left( 1 - |u(t)|^2 \Omega(t) \right)^{p} \right] .
\end{eqnarray}
The decoherence dynamics of the superposition state $|\psi_0\rangle = 1/\sqrt{2}\left( |n\rangle + |m\rangle \right)$
can be examined through the time evolution of its Wigner function, given by
\begin{eqnarray}
&~& W_{n}^{m} (z,t) = \frac{1}{2} \left[ W_{n}^{n} (z,t) + W_{m}^{m} (z,t) \right] \\
\nonumber
&+& \frac{1}{2} W_{0}^{0} (z,t) \sum_{p=0}^{\min(n,m)} \frac{\sqrt{n!}\sqrt{m!}}{p!(n-p)!(m-p)!} \times \\
\nonumber
&& \{ \left( z^{\ast} \Omega(t) u(t) \right)^{n-p} \left( z \Omega(t) u^{\ast}(t) \right)^{m-p} \left( 1 - |u(t)|^2 \Omega(t) \right)^{p} + \\
\nonumber
&& \left( z^{\ast} \Omega(t) u(t) \right)^{m-p} \left( z \Omega(t) u^{\ast}(t) \right)^{n-p} \left( 1 - |u(t)|^2 \Omega(t) \right)^{p} \} ,
\end{eqnarray}
where $W_{m}^{m} (z,t)$ is the same as $W_{n}^{n} (z,t)$ with $n$ being replaced by $m$. In figure~\ref{Super2}, the time
dependent snapshots of the Wigner functions are shown at four different times: $\omega_0 t = 0, ~1, ~1.5, ~2$.
Figure~\ref{Super2}(a) describes the decoherence dynamics of the resonator in the presence of the low frequency
$1/f^x$ noise with $x=0.25$ at a finite temperature $T=2.5$~K, and the resonator is initially prepared
in a superposition state $1/\sqrt{2}\left( |0\rangle + |3\rangle \right)$. The interference pattern consisting
of three positive and three negative peaks are caused by the superposition between the states $|0\rangle$
and $|3\rangle$. As time evolves, the off-diagonal elements of the density matrix decays with time and the
positive and negative peaks disappear. In figure~\ref{Super2}(b), we plot the time evolution of the Wigner function
for the same superposition state $1/\sqrt{2}\left( |0\rangle + |3\rangle \right)$ in presence of the $1/f^x$ noise
with $x=0.9999$. We see distinct decoherence dynamics for the resonator initial state $1/\sqrt{2}\left( |0\rangle + |3\rangle \right)$
under $1/f^x$ noise with different ``$x$'' values, where the decoherence rate is much faster for $x=0.9999$.
Next, we plot (figures~\ref{Super2}(c) and (d)) the decoherence dynamics of the resonator state $1/\sqrt{2}\left( |2\rangle + |3\rangle \right)$,
where the Wigner distribution functions show different decay dynamics of the interference fringe pattern due to the low
frequency $1/f^x$ noise (with $x=0.25$ and $x=0.9999$) of the bosonic reservoir at finite temperature $T=2.5$~K.
Faster decoherence dynamics is again observed as we increase the $x$ value of the noise spectrum.

\section{Conclusion}\label{sec:conclusion}
In conclusion, the exact decoherence dynamics of a quantum resonator coupled to a low frequency
bosonic reservoir is explored. The noise power spectrum is calculated analytically and exactly using the exact solutions
of the nonequilibrium Green's functions. It is found that the $1/f^x$ power law behavior of the noise
spectrum is valid for a very narrow range of $\eta$ and $\omega$ when both the frequency and the
coupling strengths are very small, in comparison with the energy scale of the system. The non-Markovian dynamics
of the resonator in the weak coupling regime is produced by the noise effect. The correlation Green's function $v(t)$
and hence the fluctuation coefficient $\widetilde{\gamma}(t)$ shows a long-time non-Markovian oscillatory behavior
which is qualitatively different from the Markovian dissipation dynamics described by the propagating Green's function
$u(t)$ and the dissipation coefficient $\gamma(t)$ in the ultra-weak coupling regime, in particular, in the $1/f$ regime.
We have shown through the exact master equation the evolution of a number of nonclassical photon states of the resonator
in the presence of $1/f$ noise, where the finite temperature effect of the bosonic reservoir is also examined. The faster
decoherence behavior due to the $1/f^{x}$ noise is demonstrated by increasing the $x$ value. Our analysis is also applicable
to another nano/micro mechanical oscillators, and we believe that the results presented here can enhance the understanding
of non-Markovian decoherence dynamics for many solid state quantum devices in the {\it very weak} system-reservoir
coupling regime, when the $1/f$ noise is dominated.

\begin{acknowledgments}
This work is supported by the National Science Council of ROC under Contract
No. NSC-102-2112-M-006-016-MY3 and The National Center for Theoretical Sciences.
It is also supported in part by the Headquarters
of University Advancement at the National Cheng Kung University, which is sponsored
by the Ministry of Education, Taiwan, ROC.
\end{acknowledgments}

\end{document}